\pdfoutput=1

\documentclass[twocolumn,showpacs]{revtex4}
\usepackage{amsmath,amssymb}
\usepackage{graphicx,color,hyperref}
\hypersetup{  
  pdffitwindow=false, pdfstartview={FitH}, urlbordercolor={1 1 1}  
}

\begin{document}

\title{ \LARGE
  Conditions for Conductance Quantization in Mesoscopic\\ 
  Dirac Systems on the Examples of Graphene\\ 
  Nanoconstrictions\\ 
}

\author{Grzegorz Rut}
\affiliation{Marian Smoluchowski Institute of Physics, 
Jagiellonian University, Reymonta 4, PL--30059 Krak\'{o}w, Poland}
\author{Adam Rycerz}
\affiliation{Marian Smoluchowski Institute of Physics, 
Jagiellonian University, Reymonta 4, PL--30059 Krak\'{o}w, Poland}

\begin{abstract}
Ballistic transport through an impurity-free section of the Corbino
disk in graphene is investigated by means of the Landauer-B\"{u}ttiker
formalism in the mesoscopic limit. In the linear-response regime 
the conductance is quantized in steps close to integer multiples of 
$4e^{2}/h$, yet Fabry-Perot oscillations are strongly suppressed. 
The quantization arises for small opening angles
$\theta\lesssim\pi/3$ and large radii ratios $R_2/R_1\gtrsim{}10$. 
We find that the condition for emergence of the $n$-th conductance step  
can be written as $\sqrt{n}\theta/\pi\ll1$.
A brief comparison with the conductance spectra of graphene nanoribbons
with parallel edges is also provided.
\end{abstract}

\date{October 5, 2014}
\maketitle

\section{Introduction}

Conductance quantization was observed a quarter-century ago in heterostructures
with two-dimensional electron gas (2DEG) \cite{key-1}. The
emergence of quantization steps as multiples of $2e^{2}/h$ was swiftly
associated to finite number of transmission modes. Further theoretical
investigation revealed the generic conditions under which 
conductance quantization appears in systems with constrictions 
\cite{key-3,key-4}.
It is predicted that conductance of Corbino disks in 2DEG is also quantized, 
yet in odd-integer multiples of $2e^{2}/h$ \cite{key-5}. Unfortunately,
the experimental confirmation of this result is missing so far.

In the case of graphene, theoretical calculations predict the emergence
of conductance quantization in multiples of $4e^{2}/h$ for nanoribbons
(GNRs) as well as for systems with modulated width 
\cite{key-6,Ryc07,Wur09,key-7}.
Experimental demonstration of these phenomena is challenging,
mainly due to the role of disorder and boundary effects \cite{key-8}.
These issues encourage us to study other systems exhibiting conductance
quantization which may be more resistant to the above-mentioned
factors. 

\begin{figure}[!b]
  \centerline{\includegraphics[width=0.5\linewidth]{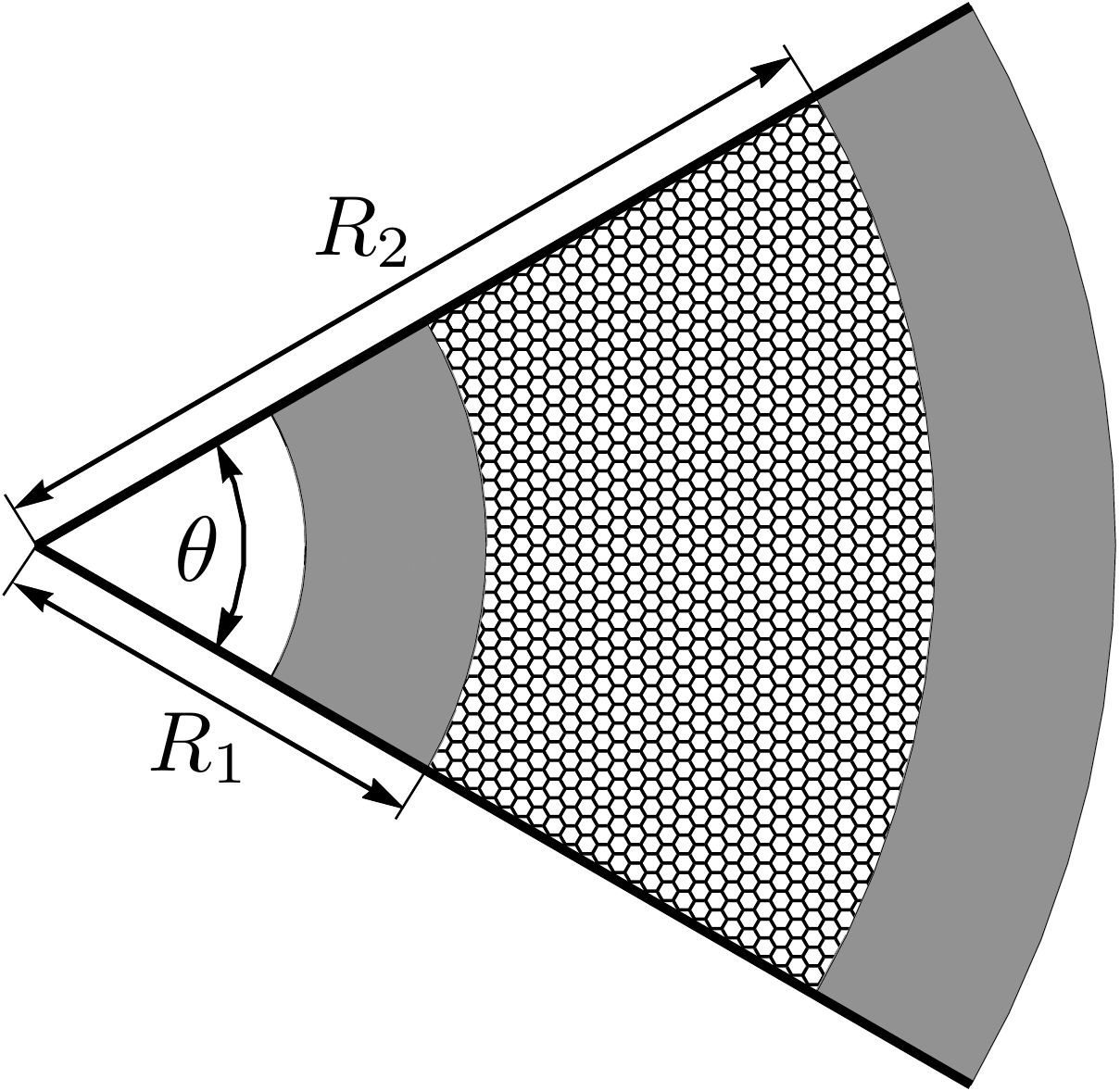}}
  \caption{\label{fig:Disk-section}
    A section of the Corbino disk in graphene attached to two metal contacts 
    (shaded areas). Tick lines at the system edges depict infinite-mass 
    boundary conditions. The opening angle $\theta=\pi/3$ and the radii ratio 
    $R_2/R_1=2$ are set for an illustration only. 
  }
\end{figure}

Transport properties of the full Corbino disk in graphene were discussed
by numerous authors \cite{key-9,Ryc10,Kha13}. 
In contrast to a~similar disk in 2DEG \cite{key-5}, 
conductance of the graphene-based system is not quantized. 
In the case of finite disk sections,
systems with wide opening angles $\theta$ (see Fig.\ref{fig:Disk-section})
should exhibit a behavior similar to complete disks as currents at
the edges play a minor role. On the other hand, narrow section strongly
resemble GNR, and thus one could raise a question: {\em At which opening
angle the quantization will emerge?} In this paper we show that conductance
steps may appear for disk sections, provided that the ratio of outer to inner
radius $R_2/R_1$ is large, and the opening angle is narrow. 

The paper is organized as follows: In Sec.\ II we discuss solutions
of the Dirac equation for a system with cylindrical symmetry. Following
Berry and Mondragon \cite{key-10}, we then impose the so-called infinite-mass 
boundary conditions \cite{Akh08}. 
In Sec.\ III we discuss the exact results of mode-matching for various 
radii ratios and opening angles.
In Sec.\ IV, the semiclassical approximation for transmission probability
is used to determine the conditions for conductance quantization in 
mesoscopic Dirac systems. For such systems, the step width is 
$\propto\sqrt{n}$ (where $n$ is the channel index), 
thus steps corresponding to large $n$ are smeared out. 
Also in Sec.\ IV, the conductance spectra a~disk section and GNR are
compared. 

\section{Model}
Our system is a section of the Corbino disk in graphene characterized
by the opening angle $\theta$ and the inner (outer) radius $R_{1}$
($R_{2}$) (see Fig.\ref{fig:Disk-section}). The leads are modelled
with heavily-doped graphene areas \cite{key-6}. 
Mode-matching analysis (see Appendices~A and B) gives the transmission 
amplitudes for quasiparticles passing through the sample area. 
The conductance is obtained by summing the transmission probabilities over 
the modes in the Landauer-B\"{u}ttiker formula 
\begin{equation}
  \label{eq:landauer}
  G=G_{0}\sum_{j}\left|t_{j}\right|^{2},
\end{equation}
with $G_{0}=4e^{2}/h$ due to spin and valley degeneracies. 

As the wavefunctions should in general posses cylindrical symmetry,
we start from the analysis of the full disk. 
The Dirac equation in polar coordinates $(r,\phi)$ can be written as
\begin{equation}
  \label{eq:diaceq}
  \left[\begin{array}{cc}
      \epsilon & 
      e^{-i\phi}\left(i\partial_{r}\!+\!\cfrac{\partial_{\phi}}{r}\right)\\
      e^{i\phi}\left(i\partial_{r}\!-\!\cfrac{\partial_{\phi}}{r}\right) & 
      \epsilon
    \end{array}\right]\left(\begin{array}{c}
      \psi_{A}\\
      \psi_{B}
    \end{array}\right)=0,
\end{equation}
where $\epsilon=\left(E-V\right)/\hbar v_{F}$, $v_{F}\approx c/300$
is the Fermi velocity, and the electrostatic potential energy is
\begin{equation}
V(r)=\begin{cases}
-V_{\infty} & \mbox{if \ensuremath{r<R_{1}\mbox{ or }r>R_{2}},}\\
0 & \mbox{if \ensuremath{R_{1}<r<R_{2}}}.
\end{cases}\label{eq:potential}
\end{equation}
Since the Hamiltonian commutes with the total angular momentum
operator $J_{z}=-i\hbar\partial_{\phi}+\hbar\sigma_{z}/2$, the wavefunction
\begin{equation}
  \label{fullpsij}
  \psi_j\left(r,\phi\right)=\exp\left[i\phi\left(j-1/2\right)\right]\left(\begin{array}{c}
      \varphi_{A}\left(r\right)\\
      \exp\left(i\phi\right)\varphi_{B}\left(r\right)
    \end{array}\right),
\end{equation}
where $j=\pm\frac{1}{2},\pm\frac{3}{2},\dots$ is the angular momentum quantum number.
Substituting $\psi_j$ into Eq. (\ref{eq:diaceq}) we can derive
\begin{equation}
  \label{varphij}
  \varphi_{j}\left(r\right)\equiv\left(\begin{array}{c}
      \varphi_{A}\left(r\right)\\
      \varphi_{B}\left(r\right)
    \end{array}\right)=\left(\begin{array}{c}
      \mbox{H}_{j-1/2}^{\left(\zeta\right)}\left(\epsilon r\right)\\
      i\mbox{H}_{j+1/2}^{\left(\zeta\right)}\left(\epsilon r\right)
    \end{array}\right),
\end{equation}
where $H_\nu^{(\zeta)}$, with $\zeta=2(1)$ for the incoming (outgoing) waves, 
is the Hankel function of the second (first) kind \cite{key-11}.
The momentum-independent radial current density is $\left({\bf j}\right)_{r}=-ev_{F}\,\psi_j^\dagger\left(\sigma_{x}\cos\phi+\sigma_{y}\sin\phi\right)\psi_j=4\lambda_{\zeta}ev_{F}/(\pi\epsilon{}r)$, with $\lambda_\zeta=(-1)^{\zeta}$. 
In the high-doping limit $\varphi_j\left(r\right)$ (\ref{varphij}) 
simplifies to 
\begin{equation}
  \label{phiasym}
  \varphi_{j}\left(r\right)\overset{\left|\epsilon\right|\rightarrow\infty}{\simeq}\sqrt{\frac{2}{\pi\epsilon r}}\exp\left[-i\lambda_\zeta\left(\epsilon r-\pi{}j/2\right)\right]\left(\begin{array}{c}
1\\
-\lambda_\zeta
\end{array}\right).
\end{equation}

Now, the sample edges are introduced to our analysis via the infinite-mass 
boundary conditions. Following Ref.\ \cite{key-10}, we demand that the angular
current vanishes at the sample edges; i.e., 
$\left({\bf j}\right)_n=\hat{\bf n}\cdot\left[\psi_j^\dagger\left(\hat{x}\sigma_{x}+\hat{y}\sigma_{y}\right)\psi_j\right]=0$, where $\hat{\bf n}$ denotes the unit vector normal to the boundary. This leads to
\begin{equation}
  \label{eq:berry}
  \psi_{B}/\psi_{A}=i\exp(\gamma),
\end{equation}
where $\gamma=0$ for $\phi=\pi/2$ or $\gamma=\theta+\pi$ for 
$\phi=\theta+\pi/2$ (without loss of generality we set the boarders at $\phi=\pi/2$ and $\phi=\theta+\pi/2$). In particular, for $\theta=\pi/(2k+1)$ with $k=0,1,2,\dots$, the solutions can be found as linear combinations of the form $a_j\psi_j+b_j\psi_{-j}$ and are given explicitly in Appendix~A. Due to Eq.\ (\ref{eq:berry}), the values of $j$ contributing to the sum in Eq.\ (\ref{eq:landauer}) are further restricted to
\begin{equation}
  \label{jvals}
  j=-\frac{\pi\left(2n-1\right)}{2\theta}, \ \ \ \ n=1,2,3,\dots.
\end{equation}
 
\begin{figure}[!t]
  \centerline{\includegraphics[width=0.9\linewidth]{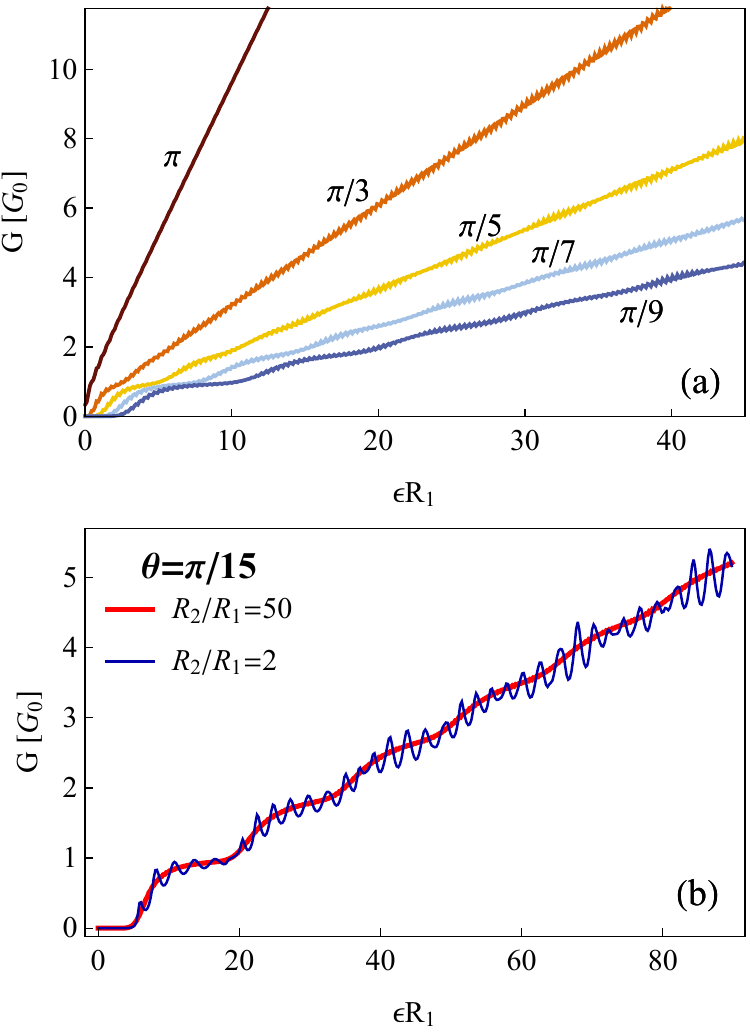}}
  \caption{\label{fig:sec}
    (a) Conductance of the disk section as a function of doping for 
    the opening angle varying from $\theta=\pi$ down to $\pi/9$ 
    (specified for each curve) and the radii ratio fixed at $R_{2}/R_{1}=10$.
    (b) Same as (a) but for $\theta=\pi/15$ and two values of $R_{2}/R_{1}$.
    Notice the suppression of the Fabry-Perot oscillations for $R_2/R_1=50$. 
  }
\end{figure}

\section{Conductance quantization}
The numerical results for disk sections with different geometric 
parameters are presented in Fig.\ \ref{fig:sec}. 
For small radii ratios $R_{2}/R_{1}\lesssim{}2$ and large opening angles 
$\theta\gtrsim\pi/3$, the approximating formula for the pseudodiffusive limit 
\cite{key-9}
\begin{equation}
G_{{\rm diff}}\approx\frac{4e^{2}}{\pi h}\,\frac{\theta}{\ln(R_{2}/R_{1})}
\end{equation}
reproduces the exact values obtained via Eq.\ (\ref{eq:landauer}) for 
$\epsilon\rightarrow{}0$.
In other cases, the conductance near the Dirac point is highly suppressed
due to the limited number of transmission modes. 
At higher dopings and for $R_{2}/R_{1}\lesssim{}10$,
we notice the Fabry-Perot oscillations arising from strong interference
between the incoming and outgoing waves in the sample area. The conductance 
quantization is clearly visible for $\theta\lesssim\pi/3$. Decreasing $\theta$, 
one can systematically increase the number of sharp conductance steps 
(see Fig.~\ref{fig:sec}a). 

\begin{figure}[!t]
  \centerline{\includegraphics[width=0.9\linewidth]{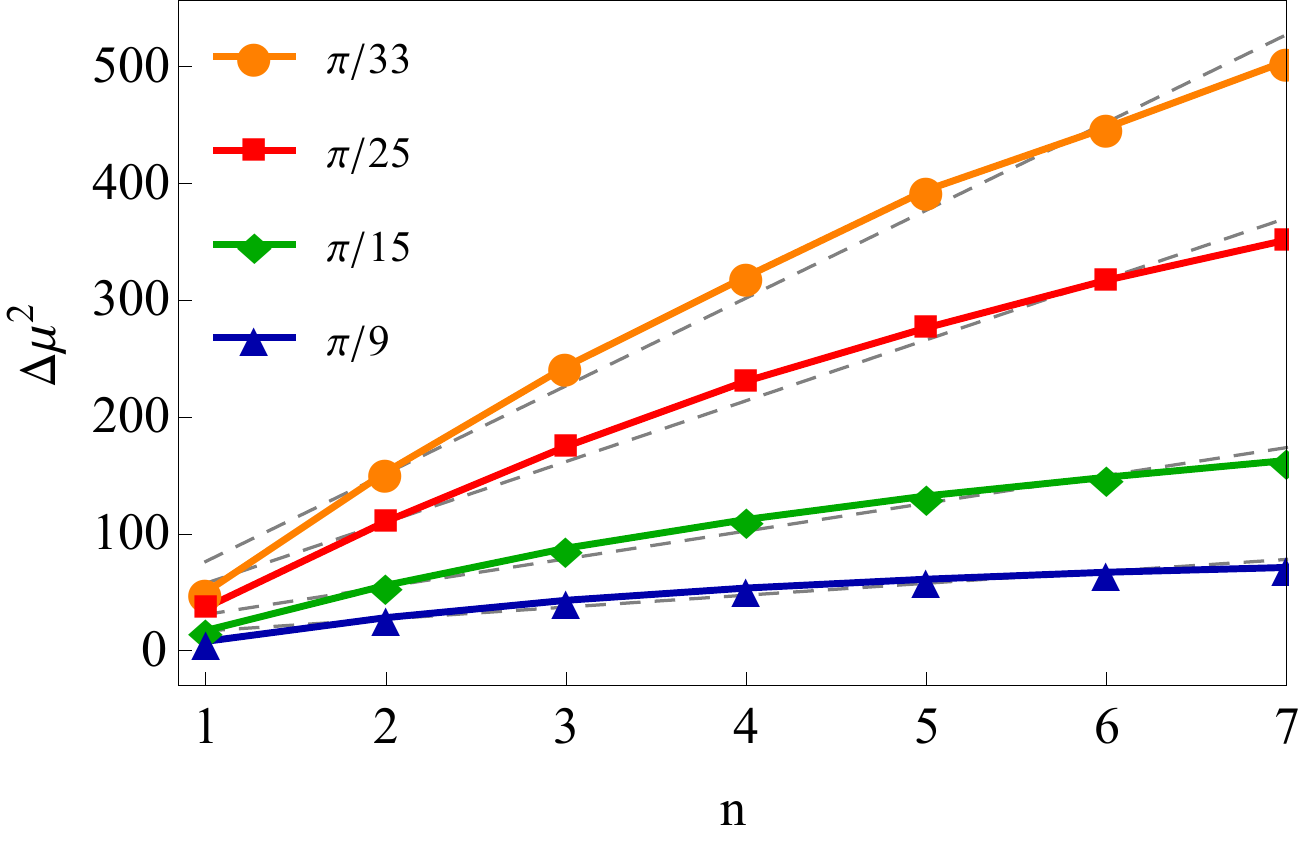}}
  \caption{\label{step}
    Squared width $\Delta\mu^{2}$ versus the step index $n$ for $R_2/R_1=10$
    and different values of $\theta$. Solid lines are guides for the eye only;
    dashed lines depict best-fitted linear dependence of $\Delta\mu^2$
    on $n$.
  }
\end{figure}

To describe the above-mentioned effect in a~quantitative manner, 
we plotted (in Fig.~\ref{step}) the squared step width $\Delta\mu^{2}$
of several consecutive conductance steps ($1\leqslant{}n\leqslant{}7$) 
for $R_2/R_1=10$ and different 
angles $\theta$. The $n$-th step width is 
quantified by the inverse slope of the straight line least-square fitted
to the exact conductance-doping dependence; i.e.,
\begin{equation}
  G/G_0\approx\frac{1}{\sqrt{\Delta\mu^2}}\,\epsilon{}R_1+{\rm const},
\end{equation}
where the fitting is performed near the inflection point corresponding to the
$n$-th conductance step.
Remarkably, $\Delta\mu^2$ increases systematically with $n$.
This observation can be rationalized by calculating the transmission probability
for electrostatic potential barrier within the semiclassical approximation
\cite{key-13}. For the classically forbidden regime, $R_1<r<j/\epsilon$, one 
can write
\begin{equation}
  \label{eq:trans}
  T_j\approx
  \exp\left[\,-2\intop_{R_{1}}^{j/\epsilon}\!dr\,
    \sqrt{\left( \frac{j}{r} \right)^{2}-\epsilon^{2}}\ \right],
\end{equation}
where $j/r$ [with $j$ given by Eq.\ (\ref{jvals})] plays a~role of the 
transverse wavenumber, and we have further supposed that $R_2\gg{}R_1$.
Each individual step, associated with the inflection point on the 
conductance-doping plot, corresponds to $T_j\approx{}1/2$ for a~given $j$. 
A~clear step becomes visible when $T_j$ rises fast enough with $\epsilon$, 
such that the step width is significantly smaller than distances to the 
neighboring steps. These lead to
\begin{equation}
  \sqrt{n}\,\theta/\pi\ll{}1.\label{eq:cond}
\end{equation}
In turn, for any finite $\theta$ only a~limited number of the conductance
steps near zero doping ($n\leqslant{}n_{\rm max}$) is visible, whereas the 
higher steps get smeared out. This effect has no direct analogue in
similar Schr\"{o}dinger systems. 

\begin{figure}[!t]
  \centerline{\includegraphics[width=0.9\linewidth]{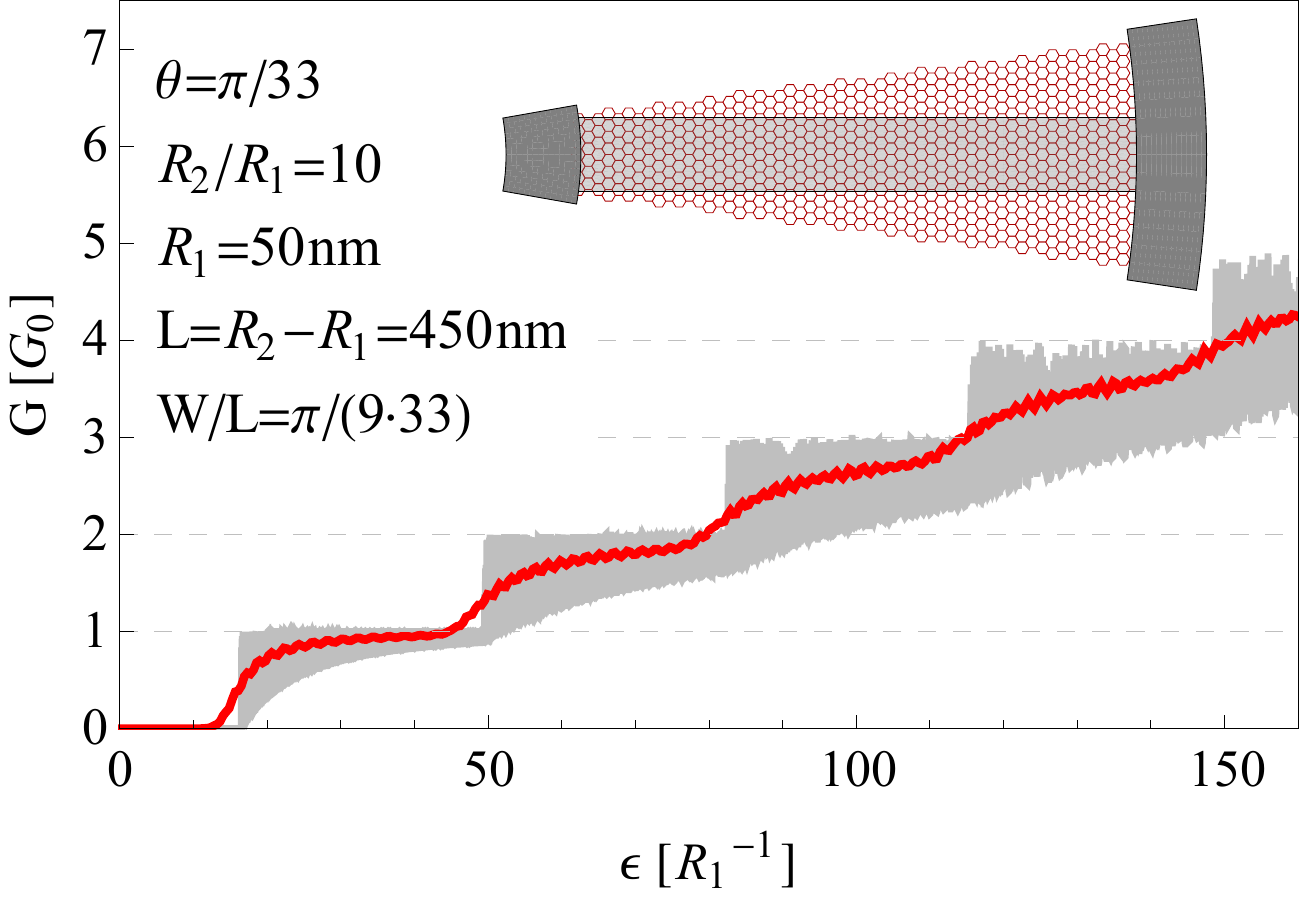}}
  \caption{\label{fig:Tworzydlo}
    Conductance as a~function of doping for graphene nanoribbon 
    (thin gray line) and narrow disk section (thick red line). 
    Inset: Schematics of the two systems considered.
  }
\end{figure}

We compare now our results with more familiar conductance quantization 
appearing for GNRs, using the analytic formula for a~strip with infinite-mass 
boundary conditions derived by Tworzyd\l{}o {\em et al.} \cite{key-6}. 
In fact, a~rectangular sample of the width $W=\theta R_{1}$
and the length $L=R_{2}-R_{1}$ essentially reproduces a~geometric quantization
appearing in a disk section for small opening angles. 
As shown in Fig.~\ref{fig:Tworzydlo}, the conductance-doping curves for the
two systems closely follow each other, except from the Fabry-Perot 
oscillations present in GNR and strongly suppressed in the disk section with 
nonparallel borders.

\section{Conclusion}
We have investigated ballistic charge transport through a~finite section of 
the Corbino disk in graphene with the infinite-mass boundaries. 
The system conductance as a~function of doping shows sharp quantization
steps for opening angles $\theta\lesssim\pi/3$.
In comparison to the situation in graphene nanoribbons, Fabry-Perot 
oscillations are strongly suppressed, particularly for large radii ratios 
$R_2/R_1\gtrsim{}10$. For these reasons, our theoretical study suggests
that a~narrow section of the disk, or a~triangle, may be the most suitable
sample geometry for experimental demonstration of the conductance quantization 
in graphene or other Dirac system. 

Additionally, a~special feature of the conductance-doping dependence
for Dirac systems has been identified. Namely, the quantization steps 
are blurred such that the step width is proportional to $\sqrt{n}$, with
$n$ being the step number. 
This observation helps to understand why only a~very 
limited number of sharp conductance steps were identified so far in both 
experimental \cite{key-8} and numerical studies \cite{key-7}.

\section*{Acknowledgements}

The work was supported by the National Science Centre of Poland (NCN)
via Grant No.\ N--N202--031440, and partly by Foundation for Polish Science
(FNP) under the program TEAM. Some computations were performed using the PL-Grid infrastructure. 

\begin{widetext}

\section*{Appendix A: Wavefunctions}
In this Appendix we give explicitly the pairs of linearly-independent 
solutions $\left[f_{A,j},f_{B,j}\right]^T$ and 
$\left[g_{A,j},g_{B,j}\right]^T$ of Eq.\ (\ref{eq:diaceq}) with the boundary 
conditions (\ref{eq:berry}). For the leads ($r<R_1$ or $r>R_2$) 
we define the dimensionless variable $\rho=\epsilon_\infty{}r$ and get
\begin{eqnarray}
f_{\alpha,j}^{L}(\rho,\phi) & = & \sqrt{\frac{8}{\pi\rho}}\exp\left[ i\left(\rho\mp\phi/2\right)\right] \cos\left[j\left(\phi-\frac{\pi}{2}\right)\right], \label{falplead} \\
g_{\alpha,j}^{L}(\rho,\phi) & = & \pm\sqrt{\frac{8}{\pi\rho}}\exp\left[ -i\left(\rho\pm\phi/2\right)\right] \cos\left[j\left(\phi+\frac{\pi}{2}\right)\right] \label{galplead}, 
\end{eqnarray}
where the upper (lower) signs correspond to the sublattice index 
$\alpha=A$ ($\alpha=B$). Similarly, for the sample area ($R_1<r<R_2$) 
$\rho=\epsilon{}r$, and the wavefunctions read
\begin{eqnarray}
f_{A,j}^{S}(\rho,\phi) & = & \exp\left[i(j+1/2)(\pi-\phi)\right]\mbox{H}_{j+1/2}^{(1)}(\rho)+\exp\left[i\left(j-1/2\right)\phi\right]\mbox{H}_{j-1/2}^{(1)}(\rho), \\
f_{B,j}^{S}(\rho,\phi) & = & i\left\{\exp\left[i\left(j+1/2\right)\phi\right]\mbox{H}_{j+1/2}^{(1)}(\rho)+\exp\left[i(j-1/2)(\pi-\phi)\right]\mbox{H}_{j-1/2}^{(1)}(\rho)\right\}, \\
g_{A,j}^{S}(\rho,\phi) & = & \exp\left[-i(j+1/2)(\phi+\pi)\right]\mbox{H}_{j+1/2}^{(2)}(\rho)+\exp\left[i\left(j-1/2\right)\phi\right]\mbox{H}_{j-1/2}^{(2)}(\rho), \\
g_{B,j}^{S}(\rho,\phi) & = & i\left\{\exp\left[i\left(j+1/2\right)\phi\right]\mbox{H}_{j+1/2}^{(2)}(\rho)+\exp\left[-i(j-1/2)(\phi+\pi)\right]\mbox{H}_{j-1/2}^{(2)}(\rho)\right\}. 
\end{eqnarray}

\section*{Appendix B: Mode-matching}
The current conservation conditions at $r=R_1$ and $r=R_2$ lead to the system 
of linear equations
\begin{equation}
\left(\begin{array}{cccc}
0 & -f_{A,j}^{L}(\epsilon_{\infty}R_{1},\phi) & f_{A,j}^{S}(\epsilon R_{1},\phi) & g_{A,j}^{S}(\epsilon R_{1},\phi) \\
0 & -f_{B,j}^{L}(\epsilon_{\infty}R_{1},\phi) & f_{B,j}^{S}(\epsilon R_{1},\phi) & g_{B,j}^{S}(\epsilon R_{1},\phi) \\
-g_{A,j}^{L}(\epsilon_{\infty}R_{2},\phi) & 0 & f_{A,j}^{S}(\epsilon R_{2},\phi) & g_{A,j}^{S}(\epsilon R_{2},\phi) \\
-g_{B,j}^{L}(\epsilon_{\infty}R_{2},\phi) & 0 & f_{B,j}^{S}(\epsilon R_{2},\phi) & g_{B,j}^{S}(\epsilon R_{2},\phi)
\end{array}\right)\left(\begin{array}{c}
t_{j}\\
r_{j}\\
a_{j}\\
b_{j}
\end{array}\right)=\left(\begin{array}{c}
g_{A,j}^{L}(\epsilon_{\infty}R_{1},\phi)\\
g_{B,j}^{L}(\epsilon_{\infty}R_{1},\phi)\\
0\\
0
\end{array}\right),
\end{equation}
where we have supposed that the wave is incident from the inner lead. 
We further notice that the transmission probability $|t_j|^2$ is insensitive
to the specific value of $\epsilon_\infty$, as it only affects the phases of 
wavefunctions $f_{\alpha,j}^{L}$ (\ref{falplead}) and $g_{\alpha,j}^{L}$ 
(\ref{galplead}). 

\end{widetext}

\end{document}